# Mode-locked ultrashort pulse generation from on-chip normal dispersion microresonators

S.-W. Huang[1,2,*], H. Zhou[1], J. Yang[1,2,], J. F. McMillan[1], A. Matsko[3], M. Yu[4], D.-L. Kwong[4], L. Maleki[3], and C. W. Wong[1, 2,†]

[1] Optical Nanostructures Laboratory, Center for Integrated Science and Engineering, Solid-State Science and Engineering, and Mechanical Engineering, Columbia University, New York, 10027

[2] Mesoscopic Optics and Quantum Electronics, University of California, Los Angeles, CA 90095

[3] OEwaves Inc, Pasadena, CA 91107

[4] Institute of Microelectronics, Singapore, Singapore 117685

**Abstract:** We describe the generation of stable mode-locked pulse trains from on-chip normal dispersion microresonators. The excitation of hyper-parametric oscillation is facilitated by the local dispersion disruptions induced by mode interactions. The system is then driven from hyper-parametric oscillation to the mode-locked state with over 200 nm spectral width by controlled pump power and detuning. With the continuous-wave driven nonlinearity, the pulses sit on a pedestal, akin to a cavity soliton. We identify the importance of pump detuning and wavelength-dependent quality factors in stabilizing and shaping the pulse structure, to achieve a single pulse inside the cavity. We examine the mode-locking dynamics by numerically solving the master equation and provide analytic solutions under appropriate approximations.

Recently continuous-wave (cw) pumped monolithic microresonators emerge as promising platforms for compact optical frequency comb generation [1–11]. With anomalous group-velocity-dispersion (GVD) and self-phase modulation (SPM), optical solitons can be generated [12,13], and remarkably broad bandwidths [6] and RF-optical stability [3] have been demonstrated. Obtaining anomalous GVD broadly across arbitrary center frequencies, however, is challenging for microresonators [14]. Dispersion engineering by conformal coating [15–17] and waveguide shaping [18] are possible, but often lead to lower quality factors ($Q$s). Alternatively, frequency comb and ultrashort pulse generation from normal GVD microresonators has been theoretically predicted [19–21] and comb-like spectra from normal GVD crystalline resonators were recently measured [22,23]. Further investigation into this normal GVD architecture, especially in the time-domain and that of coherent mode-locking, will open up new fields in chip-scale oscillators, waveform generation, and ultrafast spectroscopy.

Here we report mode-locked pulse generation from on-chip normal dispersion microresonators. The observation is supported by phase noise characterization, frequency-resolved optical gating (FROG) pulse measurement, and numerical modeling [24,25]. The phase retrieval from the FROG measurement reveals a pulse structure akin to a cavity soliton: a 74 fs mode-locked pulse sitting on a cw background. Numerical modeling of the cw-driven nonlinear microresonator, capturing the full spectra with the measured GVD and $Q$s, confirms the feasibility of mode-locked pulse generation and agrees with our measurements. We demonstrate, both experimentally and numerically, the importance of pump detuning and wavelength-dependent $Q$-factors in stabilizing and shaping the pulses generated from the normal GVD microresonators. Finally, we obtain the closed-form solution of the master equation under appropriate approximations, showing explicitly the connection between the microresonator parameters and the mode-locked pulse properties.

Figure 1a is the transmission of our $Si_3N_4$ microring resonator. Five modal families (3 TE and 2 TM) are identified from the transmission and each Lorentzian resonance is fitted to determine its frequency and $Q$-factor [26]. The frequency data is then used to evaluate the GVD. For the fundamental mode family, a loaded $Q$-factor of more than $10^6$ is achieved at 1600 nm while the $Q$-factors at the telecommunications C-band wavelengths are more than 4× lower due to residual N-H absorption [27]. For the higher order mode families, $Q$-factors are orders of magnitude smaller and thus no Kerr comb is generated from these mode families. $Q$-factors are also reduced at longer than 1625 nm due to increasing coupling loss. Therefore, the resonator has a distinct spectrally restricted area characterized with the highest $Q$-factor. As discussed later, this feature is critical for the mode-locked pulse generation in our normal GVD microresonators. Figure 1b shows the measured fundamental mode dispersion of our ring resonator, in a good agreement with our numerical modeling using a full-vectorial finite-element mode solver. Across the whole L-band, the fundamental mode features a normal GVD with local disruptions induced by mode interaction with the higher-order modes. Such change of local GVD facilitates the start of the hyper-parametric oscillation from our microresonator [26]. An example Kerr comb spectrum is shown in Figure 1c, with a spectral width spanning more than 200 nm.

The optical spectrum shows a clean mode structure with comb lines separated by single free spectral range (FSR) of the fundamental mode family, without identifiable noise peaks between comb lines (Figure 1d, inset). We investigated the Kerr comb coherence by measuring the RF amplitude noise with a scan range much larger than the cavity linewidth and by performing a cw heterodyne beat note measurement [28,29]. Both measurements confirmed the coherence of the Kerr comb. The use of RF amplitude noise as a measure of low phase noise operation has been demonstrated and widely employed [13,28,29]. With proper change of the pump power and detuning, the Kerr comb is driven into the low phase noise regime as shown in Figure 1d. The cw heterodyne beat note measurements are shown in Figure 1e. Besides the beat note of the cw laser with the pump laser, beat notes between the cw laser and different comb lines are also measured. All beat notes exhibit the same linewidth of 800 kHz, limited by coherence between the cw laser and the pump laser. Neither additional linewidth broadening of the comb lines relative to the pump nor multiple beat notes were observed, confirming the comb lines exhibit a similar level of phase noise as the pump.

We measured the pulse duration via sub-femto-joule sensitive second-harmonic-generation (SHG) non-collinear frequency-resolved optical gating (FROG) [30,31] without involvement of any optical amplification nor external bandpass filtering, to minimize pulse distortion. Careful checks were conducted to ensure no interferometric SH signal was collected in the FROG spectrogram [32]. Figure 2a is the spectrogram with 32 ps delay scan and it shows a pulse train with 8.7 ps period, the inverse of the fundamental mode family FSR (115.56 GHz). For better visualization, Fig. 2a is plotted on log-scale and the bright cw pump component is removed in the plot. Spectral interferometric fringes are clearly visible for delays longer than the pulse duration. This interference arises due to the presence of the cw background as it can also mix with the pulse, generating two temporally-separated FROG signal pulses. The fringes become sparse as the delay approaches zero and the patterns depend on the relative phase between the cw pump and the pulse [33]. Figure 2b and 2c is the spectrogram measured with a finer time resolution, 4 fs, and Figure 2c is the reconstructed spectrogram with a FROG error of 2.7%. Due to the complexity of the pulses, an iterative genetic algorithm is developed specifically to retrieve the spectrograms [26]. Figure 2d shows the retrieved pulse shape (red curve) and temporal phase profile (blue curve), with a 1.3 rad relative phase contrast observed within the pulse. The full-width-half-maximum (FWHM) pulse duration is measured at 74 fs, positively chirped from its transform-limited FWHM pulse duration of 55 fs. Due to the nature of the cw driven nonlinearity, the observed mode-locked pulse necessarily sits on a pedestal, analogous to a cavity soliton.

Figure 2e shows the measured optical intensity autocorrelation (AC) trace of the generated pulse train and the left panel of Figure 2f plots the zoom-in view. Of note, this is not an interferometric autocorrelation and thus the temporal fringes in the AC trace represent the actual oscillating structures of the pulse. Between the pulses, temporal fringes with a period of ~200 fs are clearly observed and these fringes arise from the presence of the primary comb lines, ~4.85 THz (42nd mode) away from the pump [26]. In addition, the right panel of Figure 2f shows the calculated AC traces of a stable transform-limited pulse train (black curve) and an unstable pulse train (red curve). As the instability results in the significantly increased background level of the AC trace, it shows that the instability of the generated pulse train is minimal and provides another confirmation of the stable mode-locked pulse generation [34].

To shed light on the pulse generation mechanism, we first performed numerical simulation solving the Lugiato-Lefever equation for 512 modes [25]. Experimentally-measured dispersion (Figure 1b) and wavelength-dependent $Q$ values (Figure S2), including the local dispersion disruptions, are entered into the modeling [26]. Figure 3a shows the simulation results, illustrating the emergence of the first pairs of hyper-parametric oscillation sidebands around the ±42nd modes. A good agreement with the experimental emergence result (inset) is achieved. With the proper pump power and detuning, a fundamentally mode-locked pulse train is generated as shown in Figure 3b. The modeled FWHM pulse duration is 110 fs and the relative phase contrast is 1.7 rad (positively chirped), in good agreements with the FROG measurements.

We next numerically examined idealistic nonlinear microresonators characterized by solely normal GVDs and symmetric wavelength-dependent $Q$ factors to elucidate the mode-locking physics [26]. Figure S14 shows the case with larger $D_2$ of 0.03 (or -2.7 MHz) and without wavelength dependence in $Q$ factors. A phase-locked Kerr comb can be generated, but the pulse duration is long and the shape complex. This is because, unlike in anomalous GVD microresonators, pulse broadening due to the normal GVD is not balanced by SPM and thus an additional mechanism has to be introduced to stabilize and shape the pulses. In Figure S15, we numerically introduce wavelength-dependent $Q$-factors, effectively a bandpass filter, and then clean mode-locked pulses are generated from the microresonator.

These are dark pulses and the exact pulse shapes depend on the bandpass filter bandwidth. Next, when $D_2$ is numerically set smaller at 0.003 (and closer to the experimental value), bright pulses can also be observed. Different from the case of large normal dispersion where only dark pulses exist, both bright and dark pulses are possible in the small normal dispersion case, depending on the exact combination of dispersion and bandpass filter bandwidth (Figure S16). It is even possible to generate square pulses directly with the correct sets of Q-factor profile, GVD, and detuning (Figure 3c). We note that the mode-locking mechanism has analogies, but is not identical, to the pulse generation mechanism in all-normal dispersion fiber lasers [35], a variant of additive pulse mode-locking [36].

To experimentally examine the effect of wavelength-dependent Q-factors, we then re-annealed the same microresonator at 1200°C to reduce the absorption in the shorter wavelengths such that the Q roll-off is less pronounced (Figure S2). Figure 4a shows the Kerr comb generated from the re-annealed microresonator, showing a smoother and broader spectrum than the one shown in Figure 1c. Similarly, the comb can be driven into a low phase noise state (Figure 4b). However, now without the effective narrow bandpass filter, mode-locked pulses are not observed as evidenced by the high background level (≈ 0.85) in the AC trace. A phase stable state without mode-locking is also observed in another recent study using a different microresonator platform [37].

Furthermore, we seek the closed-form solution of the master equation for the Kerr comb and pulse generation:

$$T_R \frac{\partial}{\partial T} A + \frac{i}{2}\left(\beta_{2\Sigma} + i\frac{T_c}{\Omega_f^2}\right)\frac{\partial^2}{\partial t^2} A - i\gamma |A|^2 A = -\left(\alpha + \frac{T_c}{2} + i\delta_0\right) A + i\sqrt{T_c P_{in}} e^{i\varphi_{in}} \qquad (1)$$

where $A(T,t)$ is the electric field envelope in the microresonator, $T_R$ the cavity roundtrip time, $t$ the retarded time, $T$ the slow time of the cavity, $\beta_{2\Sigma}$ the cavity GVD, $T_c$ the power coupling loss per roundtrip, $\Omega_f$ the spectral characteristics of the coupling, $\gamma$ the nonlinear coefficient, $\alpha$ the amplitude attenuation per roundtrip, $\delta_0$ the resonance detuning, and $\sqrt{P_{in}}e^{i\varphi_{in}}$ the cw pump. Here, for simplicity, we assume the intracavity bandpass filter results solely from wavelength-dependent coupling loss: $T_{coupling} \approx T_c\left[1 + \frac{(\omega_c-\omega)^2}{\Omega_f^2}\right]$, where $\omega_c$ is the frequency for maximal coupling. Assuming Gaussian input pulse and applying the variational method, the equations describing the mode-locked pulses are derived in equations (S9) [26]. Defining chirp $q$, pulse energy $E_p$, and the pulse duration $\tau$, and with $q^2 \gg \Omega_f^2 \tau^2 \gg 1$, we obtain the resulting solutions:

$$E_p \approx \frac{8\sqrt{10\pi}}{15} \frac{\beta_{2\Sigma}^{3/2} \Omega_f^2 \sqrt{\delta_0}}{T_c \gamma} \tag{2}$$

$$\tau \approx \frac{2\sqrt{5}}{3} \frac{\beta_{2\Sigma}^{3/2} \Omega_f^2}{T_c \sqrt{\delta_0}} \tag{3}$$

$$q \approx \frac{4\beta_{2\Sigma}\Omega_f^2}{3T_c} \tag{4}$$

By fitting the measured *Q*-factor (Figure S2) of the ±20 modes around $Q_{max}$ with the wavelength-dependent coupling loss profile defined above, a filter bandwidth of 2.3 THz is found. A chirp $q$ of 1.6 is then obtained after the filter bandwidth and the other measured parameters ($T_c = 0.003, \beta_{2\Sigma} = 17.14\ fs^2$) are entered into equation (4). This chirp is close to that obtained from the FROG measurement ($q = 1.5$), and the resulting calculated FWHM pulse duration (98 fs) is close to our measurements.

While the total power in the microresonator reduces as the pump detuning gets larger, equations (2) and (3) show the pulse energy actually increases and the pulse duration gets shorter. Overall, the pulse quality improves. It illustrates the active role of pump detuning: it is not simply a parameter that controls the coupled power in the microresonator, but an important physical factor that determines the pulse duration and energy distribution between the pulse and cw background. Furthermore, the closed-form solutions show that the pulse generated from a normal GVD microresonator is always chirped [equation (4)], and a narrower bandpass filter is necessary to keep the pulse short when the dispersion increases.

In summary we present the generation of mode-locked pulses from on-chip normal dispersion microresonator, supported by phase noise characterization, FROG pulse measurement, and numerical modeling with exact experimental parameters. The excitation of the hyper-parametric oscillation is facilitated by the local dispersion disruptions induced by mode interactions. Then the system is driven from the hyper-parametric oscillation to the mode-locked pulse generation by a proper change of the pump power and detuning. The phase retrieval from the FROG measurement reveals a 74 fs fundamentally mode-locked pulse sitting on a cw background. Numerical modeling of the cw-driven nonlinear microresonator, capturing the full spectra with the measured GVD and $Q$s, confirms the feasibility of mode-locked pulse generation and agrees with our measurements. We show, both experimentally and numerically, the importance of pump detuning and effective bandpass filtering in stabilizing and shaping the pulses generated from normal GVD microresonators. Finally, we present the closed-form solution of the master equation under appropriate approximations, facilitating the design of mode-locked pulse generation from microresonators.

**Acknowledgements:**

The authors acknowledge discussions with Erich P. Ippen, Zhenda Xie, and Jiangjun Zheng, a spectroscopic ellipsometer measurement and analysis at Brookhaven National Laboratory by Felice Gesuele and Tingyi Gu, respectively, and loan of the L-band EDFA and the RF spectrum analyzer from the Bergman and Shepard groups at Columbia respectively.

*swhuang@seas.ucla.edu

†cheewei.wong@ucla.edu

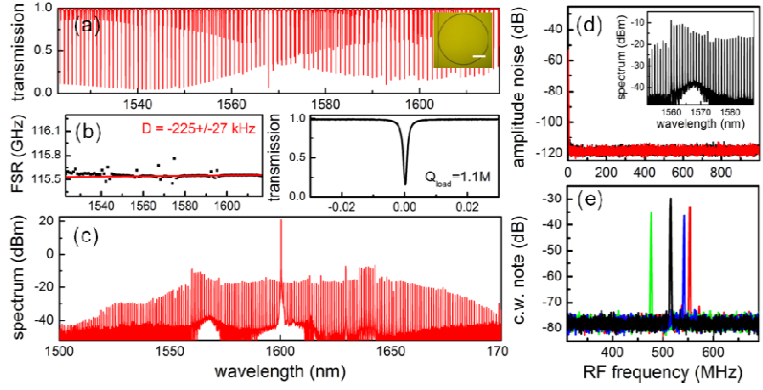

**FIG. 1** (a) Transmission of the cavity modes. Inset: an optical micrograph of the ring resonator. Scale bar: 100µm. (b) Left: wavelength dependent FSR, measuring a non-equidistance of the modes, $D_2 \equiv -\beta_2 c\omega_{FSR}^2/n_0$, of -225kHz, in a good agreement with the simulation result from a full-vector finite-element mode solver, $D_2 = -270 kHz$. Right: transmission of the cavity mode at the pump wavelength, measuring a quality factor of $1.1 \times 10^6$. (c) Example Kerr comb spectrum, with a spectral width spanning more than 200 nm. (d) RF amplitude noise of the Kerr comb (black curve) along with the detector background (red curve), indicating the low phase noise operation. Inset: a zoom-in plot of the optical spectrum, showing a clean comb structure. (e) cw heterodyne beat notes between a cw laser and different comb lines (black: pump; blue: 10th mode; red: 20th mode; green: 21st mode). No linewidth broadening of the comb lines relative to the pump is observed, showing the comb retains a similar level of phase noise as the cw laser.

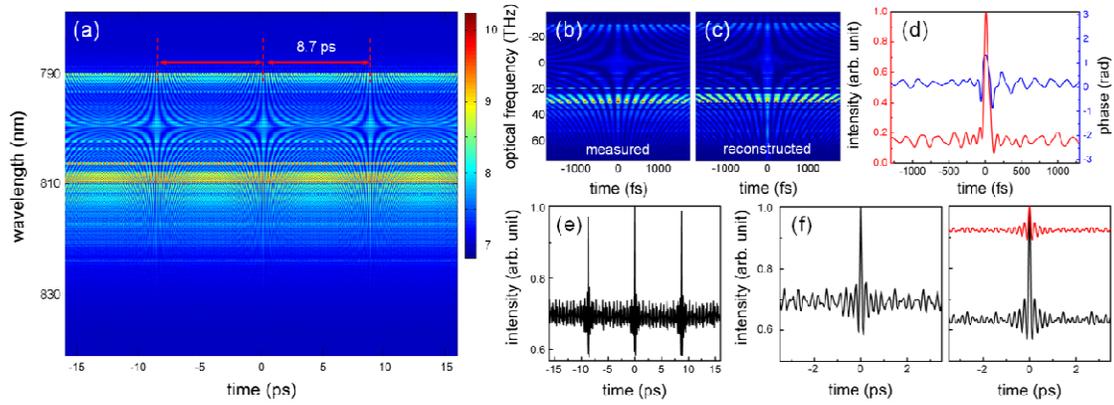

**FIG. 2** (a) FROG spectrogram with a delay scan of 32 ps, showing a fundamentally mode-locked pulse train. (b) FROG spectrogram measured with a finer time resolution of 4 fs. (c) Reconstructed FROG spectrogram achieved by use of genetic algorithms. (d) Retrieved pulse shape (red curve) and temporal phase profile (blue curve), measuring a 74 fs FWHM pulse duration. (e) Measured AC of the generated fundamentally mode-locked pulse train. (f) Left: a zoom-in plot of the measured AC. Right: the calculated ACs of a transform-limited stable pulse train (black curve) and an unstable pulse train showing a significantly larger AC background (red curve).

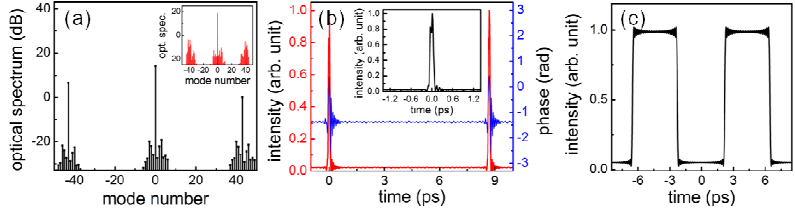

**FIG. 3** (a) Near the threshold and with a small red-detuning of 180 MHz, the first pairs of hyper-parametric oscillation sidebands emerge at around the ± 42nd modes, showing a good agreement with the experimental result (inset). (b) With the proper pump power (260 mW) and red-detuning (2.5 GHz), a mode-locked pulse train is generated. The red and blue curves are the modeled pulse shape and the temporal phase profile, respectively. Inset: a zoom-in plot of the pulse shape, showing an ultrashort FWHM pulse duration of 110 fs. (c) Square optical pulses can also be generated directly from a normal GVD microresonator. The conditions for the observation of these square pulses are $D_2 = 0.002$, red-detuning of $7.7\gamma_0$, resonance linewidth of $\gamma_j = \gamma_{j0}[1 + 0.01(j - j_0)^2]$ and pump power 25 times larger than the threshold.

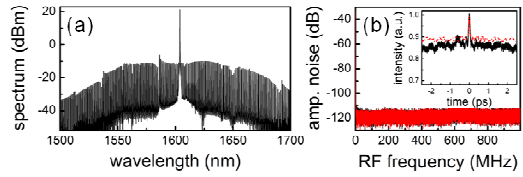

**FIG. 4** (a) Example Kerr comb spectrum from the re-annealed microresonator, showing a smoother and broader spectrum. (b) RF amplitude noise of the Kerr comb (black curve) along with the detector background (red curve). While the Kerr comb can also be driven to a low phase noise state, the high background level of the AC trace (inset) indicates the absence of mode-locked pulses. The red dashed line is the calculated AC trace assuming random spectral phases.

# Supplemental Material for

## Mode-locked pulse generation from on-chip normal dispersion microresonators

S.-W. Huang[1,*], H. Zhou[1], J. Yang[1], J. F. McMillan[1], A. Matsko[2], M. Yu[3], D.-L. Kwong[3], L. Maleki[2], and C. W. Wong[1,†]

[1] Optical Nanostructures Laboratory, Center for Integrated Science and Engineering, Solid-State Science and Engineering, and Mechanical Engineering, Columbia University, New York, 10027

[2] OEwaves Inc, Pasadena, CA 91107

[3] Institute of Microelectronics, Singapore, Singapore 117685

## I. $Si_3N_4$ ring resonator structure, refractive index and quality factor characterization

Figure S1a shows the layout of the ring resonator and the refractive index of the low pressure chemical vapor deposition (LPCVD) $Si_3N_4$. Due to the large refractive index of the $Si_3N_4$ waveguide, a 600 μm long adiabatic mode converter (the $Si_3N_4$ waveguide, embedded in the 5×5 μm$^2$ $SiO_2$ waveguide, has gradually changing widths from 0.2 μm to 1 μm) is implemented to improve the coupling efficiency from the free space to the bus waveguide. The input-output insertion loss for the waveguide does not exceed 6 dB. The refractive index was measured with an ellipsometric spectroscopy (Woollam M-2000 ellipsometer) and the red curve is the fitted Sellmeier equation assuming a single absorption resonance in the ultraviolet (Figure S1b). The fitted Sellmeier equation, $n(\lambda) = \sqrt{1 + \frac{(2.90665 \pm 0.00192)\lambda^2}{\lambda^2 - (145.05007 \pm 1.03964)^2}}$, was then imported into the COMSOL Multiphysics for the waveguide dispersion simulation, which includes both the material dispersion and the geometric dispersion.

The fabrication procedure of our microresonator: First a 3 μm thick $SiO_2$ layer was deposited via plasma-enhanced chemical vapor deposition on p-type 8" silicon wafers to serve as the under-cladding oxide. Then LPCVD was used to deposit a 725 nm silicon nitride for the ring resonators, with a gas mixture of $SiH_2Cl_2$ and $NH_3$. The resulting $Si_3N_4$ layer was patterned by optimized 248 nm deep-ultraviolet lithography and etched down to the buried $SiO_2$ via optimized reactive ion dry etching. The sidewalls were observed under SEM for an etch verticality of 88 degrees. The nitride rings were then over-cladded with a 3 μm thick SiO$_2$ layer, deposited initially with LPCVD (500 nm) and then with plasma-enhanced chemical vapor deposition (2500 nm). The device used in this study has a ring radius of 200 μm, a ring width of 2 μm, and a ring height of 0.725 μm.

Figure S2 shows the wavelength-dependent $Q$-factors of the ring resonator, determined by Lorentzian fitting of cavity resonances. The loaded $Q$ reaches its maximum (~1.4M) at 1625 nm and gradually decreases on both ends due to the residual N-H absorption at the short wavelengths and the increasing coupling loss at the long wavelengths. This effective bandpass filter plays an important role in pulse generation from our normal GVD microresonator.

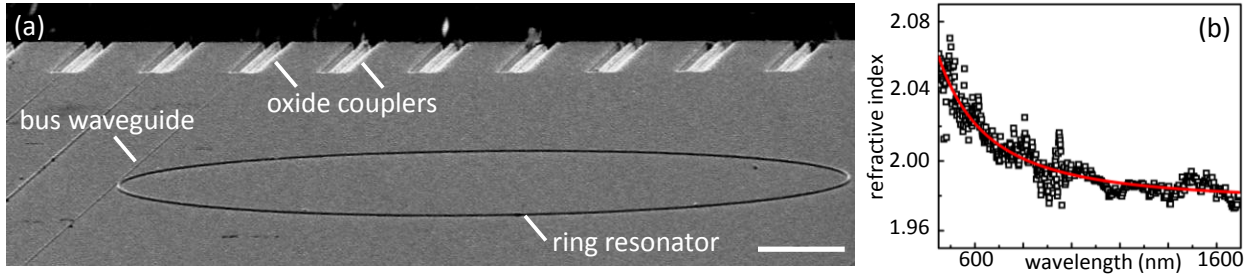

**Figure S1 | Scanning electron micrograph of the chip-scale ring resonator. a,** Layout of the ring resonator with input/output mode converters with less than 3 dB coupling loss on each facet. Scale bar: 50 μm. **b,** Spectroscopic ellipsometer measurements of the refractive index of the LPCVD Si$_3$N$_4$ for the numerical dispersion modeling.

## II. Dispersion measurement and mode interaction

Figure S3 shows the dispersions of the ring resonator calculated with a commercial full-vector finite-element-mode solver (COMSOL Multiphysics), taking into account both the waveguide dimensions and the material dispersion. Modeling is performed on 50 nm triangular spatial grid with perfectly-matched layer absorbing boundaries and 5 pm spectral resolution. Since the ring radius is large, the bending loss and the bending dispersion of the resonator waveguide are negligible in our ring resonators [S1]. The fundamental mode (TE$_{11}$) features small normal group velocity dispersion (GVD) and small third-order dispersion (TOD) across the whole telecommunication wavelength range while the first higher order mode (TE$_{21}$) possesses large anomalous GVD and large TOD. We define GVD and TOD in accordance with formulas $GVD \equiv \frac{\partial^2 \varphi}{\partial \omega^2} = \frac{\lambda^3}{2\pi c_0^2} \frac{d^2 n}{d\lambda^2}$ and $TOD \equiv \frac{\partial^3 \varphi}{\partial \omega^3} = -\frac{\lambda^4}{4\pi^2 c_0^3}\left(\lambda \frac{d^3 n}{d\lambda^3} + 3\frac{d^2 n}{d\lambda^2}\right)$.

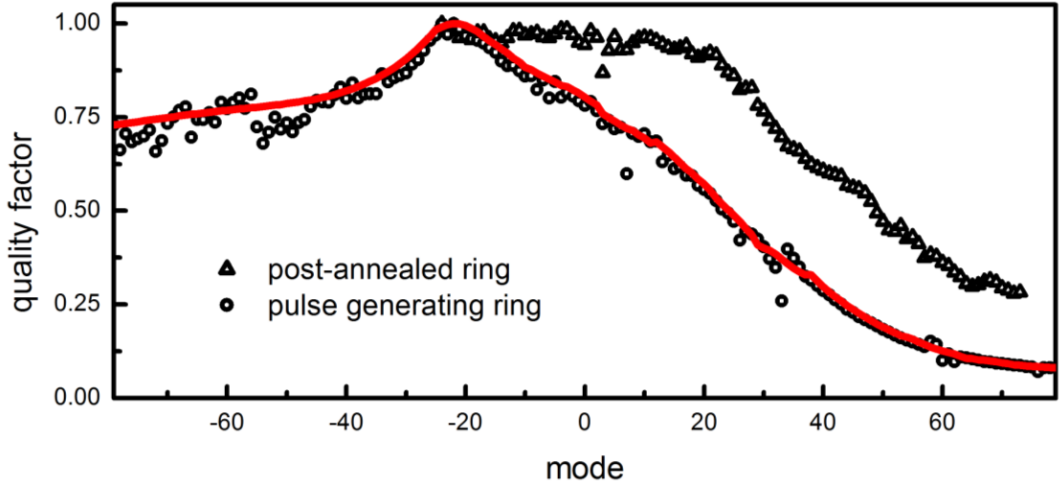

**Figure S2 | $Q$ quantification of the resonant modes.** The intrinsic absorption from the residual N-H bonds results in the loaded $Q$s' roll-off at the short wavelengths (circles). Post-annealing of the $Si_3N_4$ ring resonator lowers the concentration of the residual N-H and reduces the roll-off (triangles). At the long wavelengths, the increasing coupling loss is responsible for the $Q$ roll-off. The red curve is the fit of the loaded $Q$s, used in the numerical simulations.

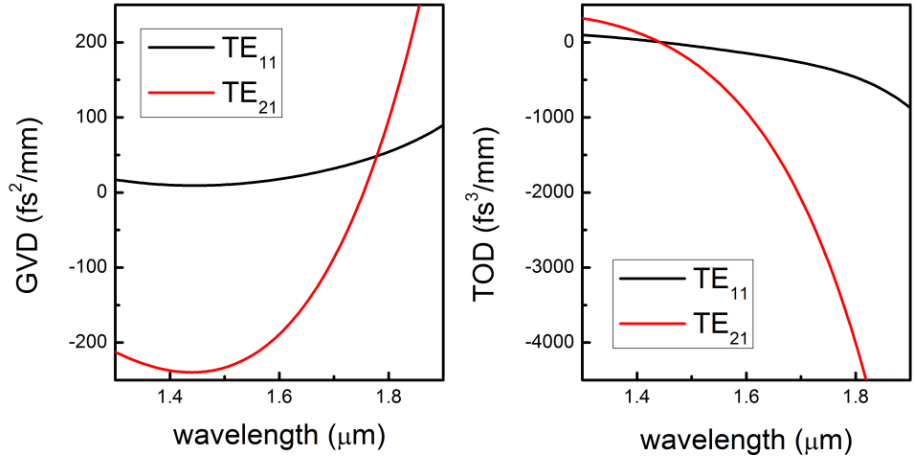

**Figure S3 | Simulated GVD and TOD of the ring resonator.** The fundamental mode features normal GVD across the whole telecommunication wavelength range while the first higher order mode possesses anomalous GVD. The fundamental mode also features small TOD at the telecommunication wavelength range, beneficial for broad comb generation.

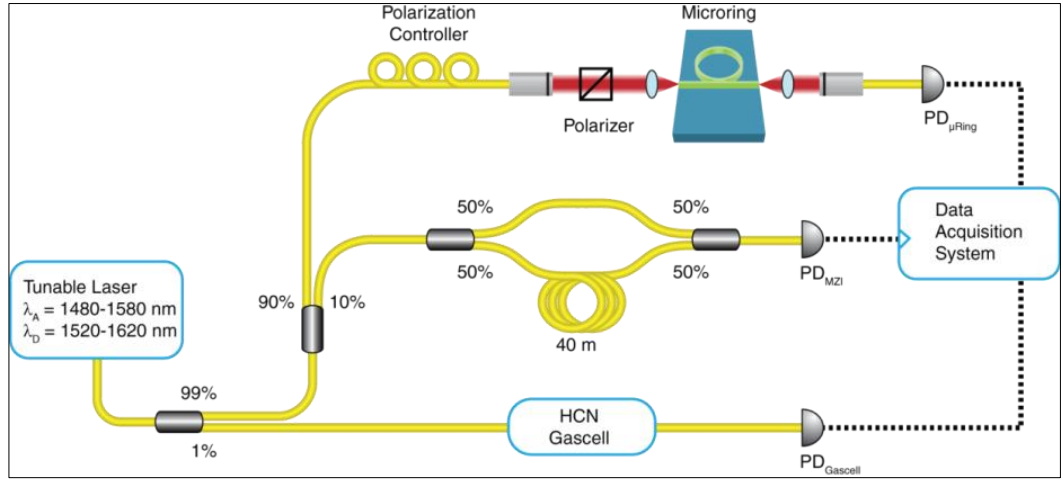

**Figure S4 | Dispersion measurement setup.** The laser is swept through its full wavelength range at 40 nm/s tuning range and the absolute wavelength is calibrated with a hydrogen cyanide gas cell. The sampling clock of the data acquisition is derived from the photodetector monitoring the laser transmission through a fiber Mach-Zehnder interferometer with 40 m unbalanced path lengths, which translates to a 5 MHz optical frequency sampling resolution. 51 absorption features of the gas cell are correlated with the wavelength sweep to determine the subsample positions.

Figure S4 shows the schematic diagram of the dispersion measurement setup. The microresonator transmission, from which quality factor and FSR values are determined, was measured using a tunable laser swept through its full wavelength tuning range at a tuning rate of 40 nm/s. For absolute wavelength calibration, 1% of the laser output was directed into a fiber coupled hydrogen cyanide gas cell (HCN-13-100, Wavelength References Inc.) and then into a photodetector ($PD_{Gascell}$). The microresonator and gas cell transmission were recorded during the laser sweep by a data acquisition system whose sample clock was derived from a photodetector ($PD_{MZI}$) monitoring the laser transmission through an unbalanced fiber Mach-Zehnder Interferometer (MZI). The MZI has a path length difference of approximately 40 m, making the measurement optical frequency sampling resolution 5 MHz. The absolute wavelength of each sweep was determined by fitting 51 absorption features present in the gas cell transmission to determine their subsample position, assigning them known traceable wavelengths [S2] and calculating a linear fit in order to determine the full sweep wavelength information. Each resonance was fitted with a Lorentzian lineshape unless a cluster of resonances were deemed too close to achieve a conclusive fit with a single Lorentzian. Then, an *N*-Lorentzian fit was utilized where *N* is the number of resonances being fitted. The dispersion of the ring resonator was then determined by analyzing the wavelength dependence of the FSR.

To compare the dispersion measurements with the COMSOL calculations, we performed two other measurements beside the one shown in Figure 1b. Figure S5a and S5b show the measured dispersions of the $TE_{21}$ mode of the microresonator used in this paper (a ring width of 2 µm) and the $TE_{11}$ mode of the microresonator with a different ring width of 1.55 µm, respectively. Both measurements show good agreements with the COMSOL calculations ($D_{mea}$ = 2.2 MHz versus $D_{sim}$ = 2.4 MHz, and $D_{mea}$ = 330 kHz v.s. $D_{sim}$ = 500 kHz). The good agreements give us confidence in the COMSOL calculations, and thus we use the calculated dispersions in the Kerr comb numerical simulation for the wavelength range not covered by the dispersion measurements (due to the unavailability of the suitable tunable laser).

It has been shown that the resonance shift due to the temperature drift is the major cause in the uncertainty of the dispersion measurements [S3]. In our measurement setup, we actively control both the ambient and the on-chip temperature and the temperature drift in 2 second is measured to be less than 5 mK. Such a temperature drift will lead to a resonance shift that can be calculated by $\frac{\Delta v}{v_0} = -\left(\alpha + \frac{1}{n}\frac{dn}{dT}\right)\Delta T$, where the thermal expansion coefficient $\alpha = 3.3 \times 10^{-6}/K$ and the thermo-optic coefficient $\frac{dn}{dT} = 2.45 \times 10^{-5}/K$ [S4, S5]. Thus the uncertainty of our dispersion measurement setup, limited by the temperature induced resonance shift, is estimated to be less than 175 kHz/mode. Experimentally, we observe measurement errors of less than ± 70 kHz/mode, in good agreement with the estimation. For each measurement, four dataset are taken and independently fit to find the dispersion. The reported *D*s are the average values and the measurement errors are the standard deviations of the four dataset. Furthermore, we also confirm the temperature drift has minimal effect when the wavelength scan speed is set higher than 20 nm/s (Figure S5c).

Of note, the non-equidistance of the modes in our ring resonator can be estimated as $D_2 = -270\ kHz$. Compared to the resonance linewidth, $2\gamma_0 = 180\ MHz$, the non-equidistance is insignificant and thus comb spacing alterations due to mode interaction are pronounced in our ring resonator [S6]. The frequency shift $\Delta_a$ of mode *a* due to interaction with mode *b* can be estimated using the formula $\Delta_a = -\frac{\kappa^2}{\Delta}$, where $\kappa$ is the interaction constant and $\Delta$ is the difference in eigenfrequencies of the interacting modes (*a* and *b*) [S6]. Even with an assumption of large $\Delta$ of 10 GHz, a small mode interaction constant $\kappa = 0.6\gamma_0$ can change the local dispersion from normal dispersion to anomalous dispersion. Similar effect was also observed and characterized in Ref. [S7].

Figure S6 plots the resonance frequency offsets with respect to the fundamental mode family (top) as well as the wavelength-dependent FSRs of the fundamental mode family (bottom). The zero crossings on the upper panel represent the wavelengths where the fundamental mode family experiences mode crossings with other higher order mode families. The lower panel then shows that the disruption of the dispersion continuity of the fundamental mode family is dominated by the mode interaction with the first higher order TE mode family.

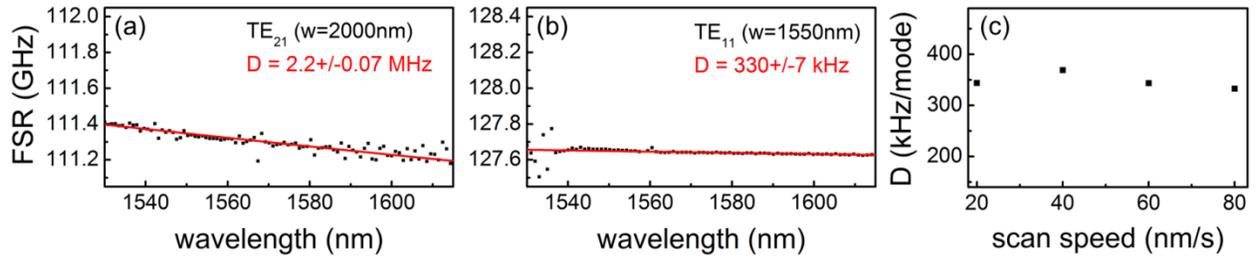

**Figure S5 | Measured dispersions.** (a) Wavelength dependence of the FSR, measuring a non-equidistance of the modes, *D*, of 2.2 MHz, in a good agreement with the COMSOL simulations, *D* = 2.4 MHz. (b) Wavelength dependence of the FSR, measuring a non-equidistance of the modes, *D*, of 330 kHz, in a good agreement with the COMSOL simulations, *D* = 500 kHz. (c) Dispersion measured at different wavelength scan speeds, showing the minimal effect of the temperature drift when the wavelength scan speed is set higher than 20 nm/s.

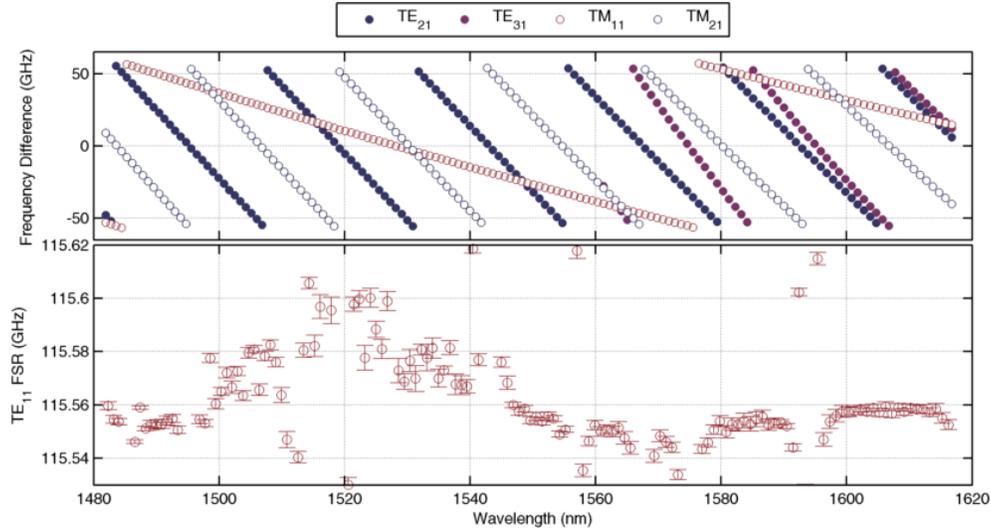

**Figure S6 | Frequency offset and FSR of the modal families.** Upper panel: The resonance frequency offsets with respect to the fundamental mode family. Lower panel: Wavelength-dependent FSRs of the fundamental mode family.

**III. Kerr comb and ultrashort pulse characterization**

Hyper-parametric oscillation in an anomalous dispersion microresonator starts from the modulation instability of the intra-cavity cw light. When the intra-cavity power exceeds a certain threshold, the cw field becomes modulated and the modes of the resonator that is phase matched start to grow. Since most materials possess positive Kerr nonlinearities, anomalous GVD is tuned in prior resonators to satisfy the phase matching condition. Increase of the optical power can result in soliton formation, leading to the generation of a broad frequency comb and short pulses.

Hyper-parametric oscillation as well as Kerr comb formation is also possible in the case of normal GVD, but a non-zero initial condition is required for frequency comb and pulse generation [S8]. In our microresonator, the comb can be ignited due to the change of local GVD resulting from the mode interaction between the fundamental mode family, which has a normal GVD, and the first higher order mode family, which has an anomalous GVD (see Figures 1b). Mode interaction enables excitation of the hyper-parametric oscillation from zero initial conditions. It is possible then to introduce a non-adiabatic change to the system parameters and transfer the system from the hyper-parametric oscillation regime to the frequency comb generation regime [S8]. Here a non-adiabatic change means a stepwise change of resonance detuning or pump power, instead of a continuous scan, in a time shorter than the time of the comb growth, which is much longer compared to the cavity lifetime [S8, S9].

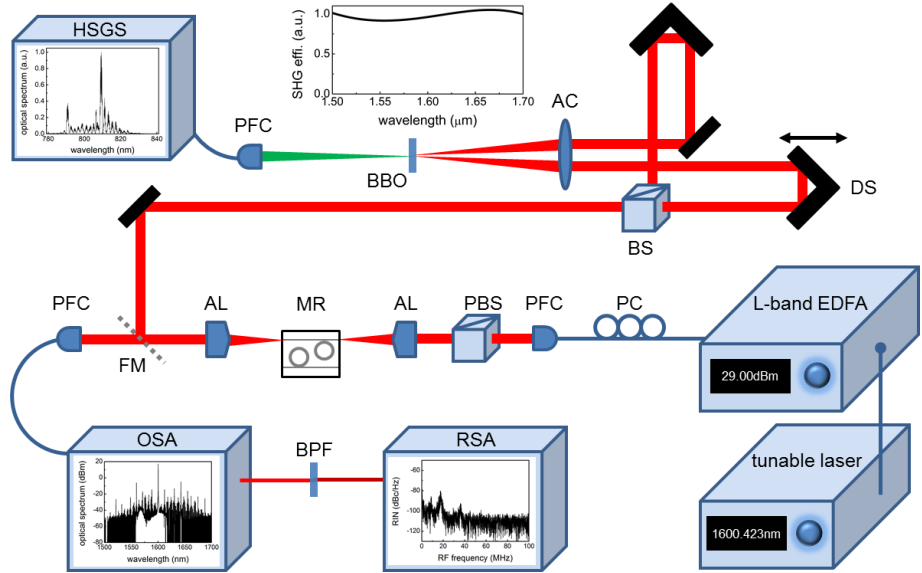

**Figure S7 | Comb characterization and FROG measurement setup.** PC, polarization controller; PFC, pigtailed fiber coupler; PBS, polarization beamsplitter; AL, aspheric lens; MR, micro-resonator; FM, flip mirror; OSA, optical spectrum analyzer; BPF, bandpass filter; RSA, RF spectrum analyzer; BS, beamsplitter; DS, delay stage; AC, achromatic lens; HSGS, high-sensitivity grating spectrometer; BBO, $\beta$-barium borate. BBO is chosen to be the second-harmonic generation crystal because it has been shown to exhibit ultrabroad phase matching bandwidth at the telecommunication wavelengths [S10, S11].

Figure S7 shows the schematic diagram of the comb and pulse generation and characterization setup. The cw pump started from an external cavity stabilized tunable laser (Santec TSL-510C). The linewidth of the laser is 200 kHz and the frequency stability over an hour is <120MHz. The pump power was increased from 8dBm to 29 dBm in an L-band EDFA (Manlight HWT-EDFA-B-SC-L30-FC/APC). A 3-paddle fiber polarization controller and a polarization beam splitter cube were used to ensure the proper coupling of TE polarization into the microresonator. The total fiber-chip-fiber loss is 6 dB. The microresonator chip was mounted on a temperature controlled stage set to 60$^\circ$C. The temperature stability over an hour is <0.1$^\circ$C so that the change in coupling loss is negligible (<0.5%). The output light was sent to an optical spectrum analyzer (Advantest Q8384) and a photodiode (Thorlabs DET01CFC) connected to an RF spectrum analyzer (Agilent E4440A) for monitoring of comb spectrum and RF amplitude noise, respectively. For the RF amplitude noise measurement, a 10 nm portion of the optical spectrum (1560 nm to 1570 nm) was filtered from the comb. The output light can also be sent by a flip mirror to the FROG setup for pulse characterization. The FROG apparatus consists of a lab-built interferometer with a 1 mm thick β-BBO crystal and a high-sensitivity grating spectrometer with a cryogenically-cooled deep-depletion 1024×256 Si CCD array (Horiba Jobin Yvon CCD-1024×256-BIDD-1LS). The FROG setup is configured in a non-collinear geometry and careful checks were done before measurements to ensure only background-free SH signals were collected. The use of dispersive optics is minimized and no fiber is used in the FROG apparatus such that the additional dispersion introduced to the pulse is only -50 fs$^2$. The FROG can detect pulses with a bandwidth of >200 nm [S10, S11] and a pulse energy of <100 aJ (10 μW average power) with a 1 second exposure time. With the sensitive FROG, no additional optical bandpass filtering and amplification is needed (minimizing pulse distortion), though there is a small amount of dispersive filtering and intensity modification with the coupling optics and ring-waveguide coupling. The FROG reconstruction was done iteratively using genetic algorithm [S12]. Genetic algorithm is a global search method based on ideas taken from evolution and is less susceptible to becoming trapped by local extrema in the search space. Both the spectral amplitudes and phases are encoded as strings of 8-bit chromosomes and two genetic operators, crossover and mutation, are used to generate the next-generation solutions. Tournament selection with elitism is employed to ensure monotonically convergence of the solution [S13]. The FROG error is defined as $\varepsilon = \sqrt{\frac{1}{N^2}\sum_{i,j=1}^{N}|S_{mea}(\omega,t) - S_{ret}(\omega,t)|^2}$, where $S_{mea}(\omega,t)$ and $S_{ret}(\omega,t)$ are the measured and reconstructed spectrograms.

To make sure the retrieval program converges accurately to the right answers, we tested the retrieval program with a few sets of simulated FROG spectrograms. Figure S8 shows a retrieval example of an ideal Gaussian pulse with dispersion randomly generated by the simulation program. We next placed the ideal Gaussian pulse with dispersion on a cw background, with results on the correct retrieval shown in Figure S9. Of note, the fringe patterns in the spectral domain have started to appear in this case. For the pulse on a cw background and for delays longer than the pulse duration, the FROG signal has two temporally-separated pulses due to the mixing between the cw background and the pulse. Such two pulses result in the spectral interference patterns. Finally, we placed the simulated pulse with dispersion on a cw background, now with additive white noise to mimic the experimental data. The amplitude of the additive white noise is chosen such that the resulting FROG errors are between 2.6% to 2.8%, close to the experimentally obtained value (2.7%). Four example retrievals are shown in the Figure S10. We emphasize that in all cases, the pulse shapes are still faithfully reconstructed and the temporal phase profiles only show from minor deviations from run-to-run.

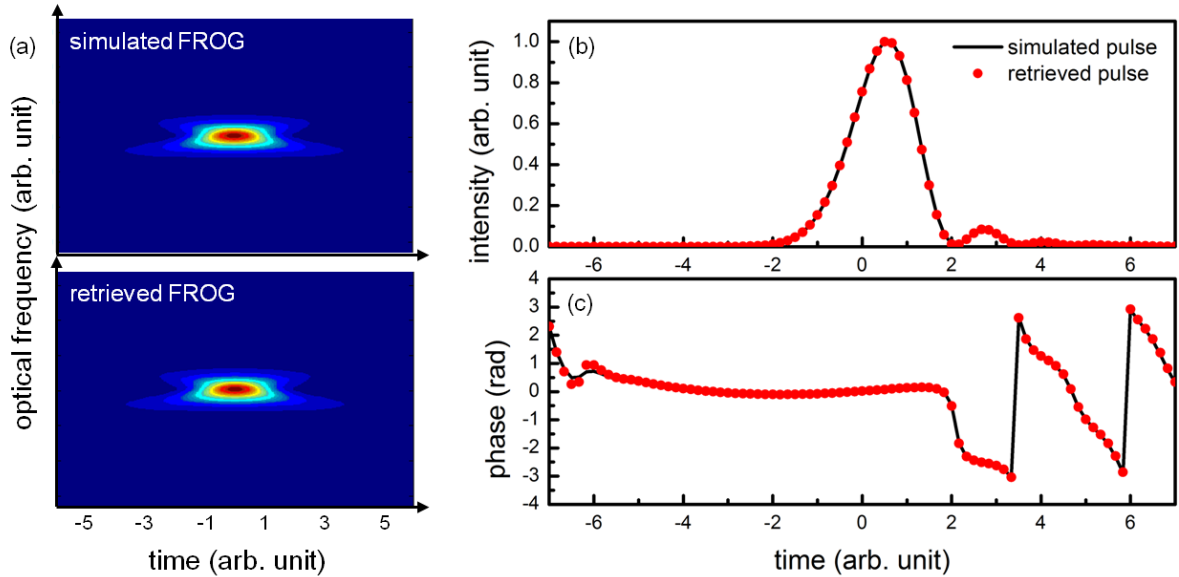

**Figure S8 | Retrieval of ideal Gaussian pulse with dispersion. a,** simulated and retrieved FROG spectrograms. **b,** intensity profile comparisons between simulated input pulse and retrieved pulse. **c,** phase profile comparisons between simulated input pulse and retrieved pulse.

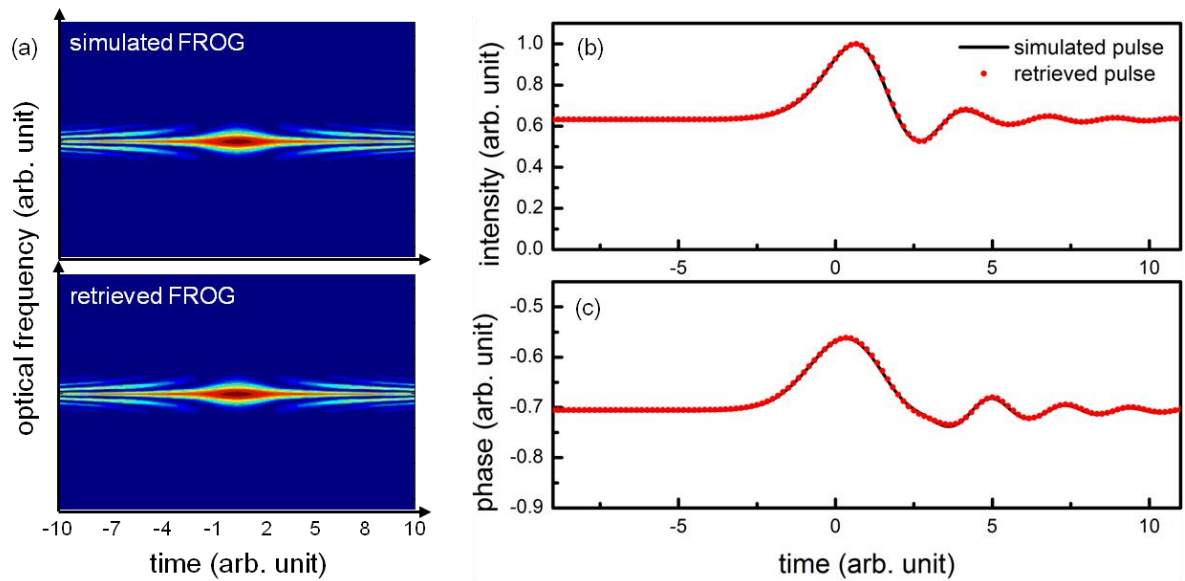

**Figure S9 | Retrieval of ideal Gaussian pulse with dispersion, on a cw background. a,** simulated and retrieved FROG spectrograms. **b,** intensity profile comparisons between simulated input pulse and retrieved pulse. **c,** phase profile comparisons between simulated input pulse and retrieved pulse.

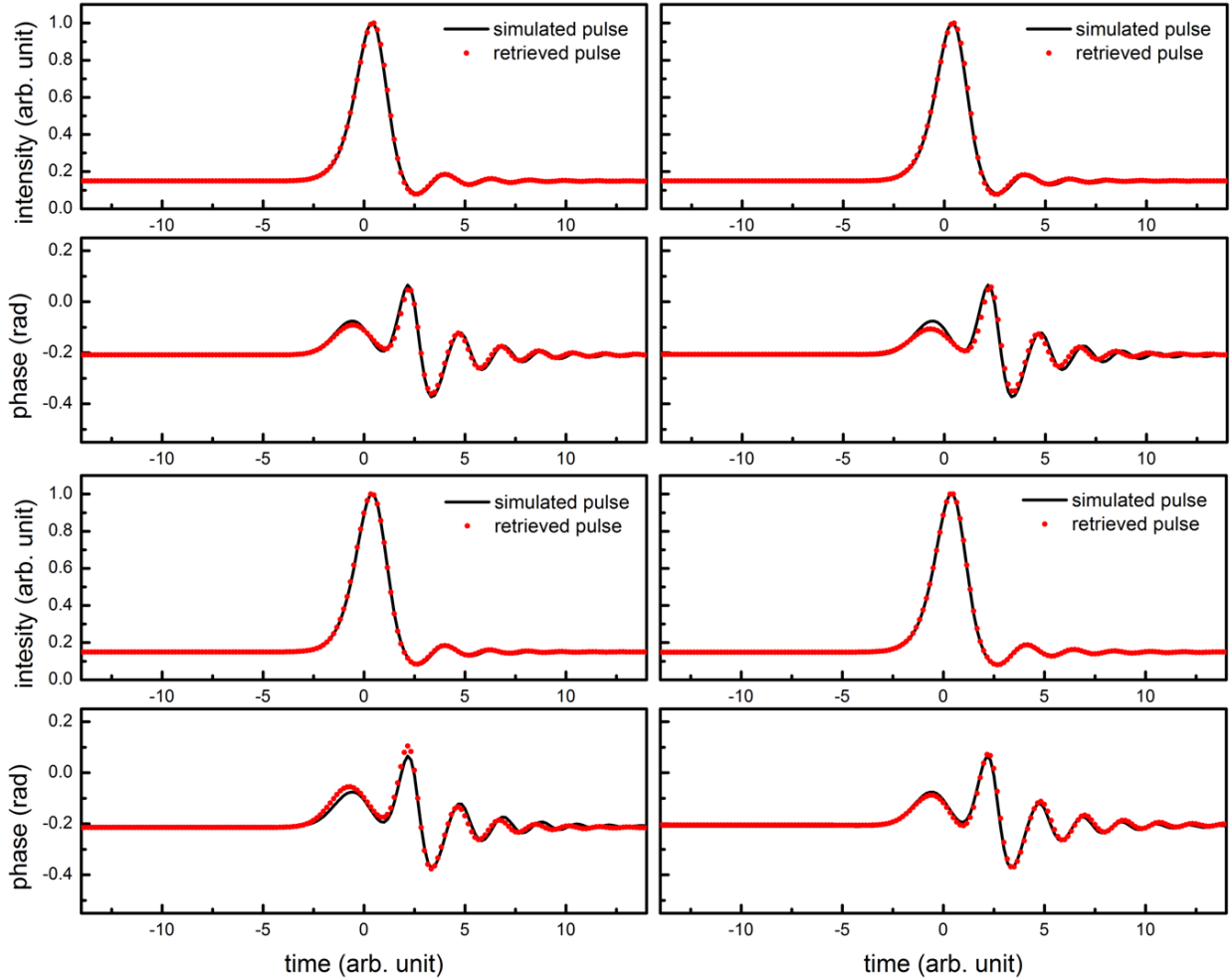

**Figure S10 | Retrievals of ideal Gaussian pulse with dispersion, on a cw background, with additive white noise.** Four example cases are illustrated. In each, the pulse shapes are faithfully reconstructed and the temporal phase profiles only show minor deviations from run-to-run.

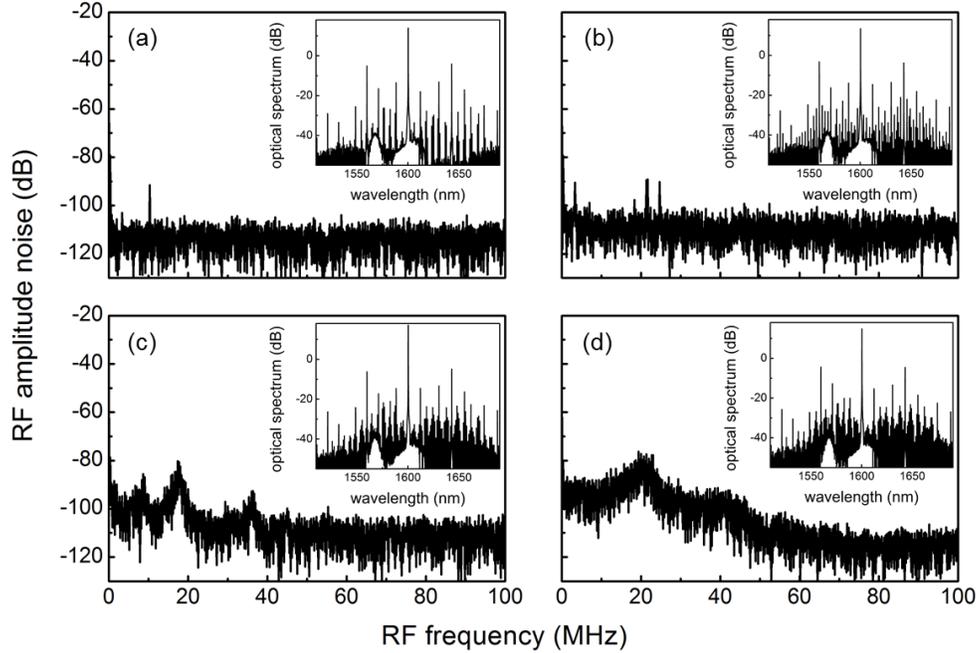

**Figure S11 | Normal dispersion Kerr comb evolution.** Growth of the RF amplitude noise and the comb spectrum (inset) are measured at four different pump detunings over a 30 pm range as the pump is tuned into the cavity resonance (from (a) to (d)). As we tune the pump wavelength further into resonance and more power is coupled into the microresonator, the bandwidth of the secondary comb families grows and the spectral overlap between them becomes more extensive, resulting in an increase of RF amplitude noise and merging of multiple RF spikes to form a continuous RF noise spectrum. After sweeping the detuning and power levels to generate a broad comb spectrum, we next perform an abrupt discrete step-jump in both detuning and power to achieve the low phase noise state, and are able to find a set of parameters at which the RF amplitude noise drops by orders of magnitude and approaches the detector background noise (Figure 1d). The phase-locked comb typically stabilizes for more than three hours.

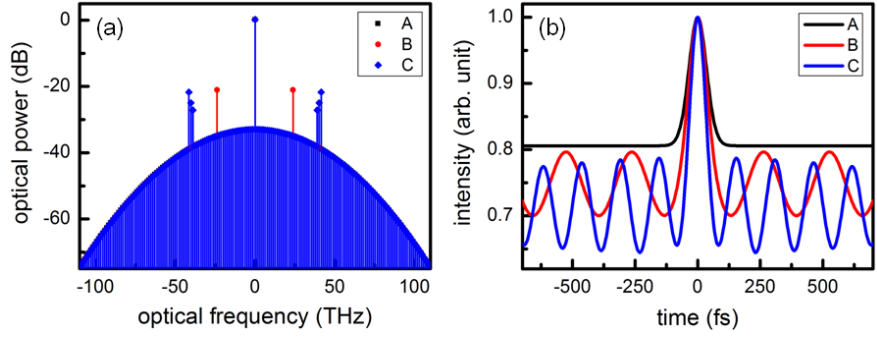

**Figure S12 | Temporal fringes resulting from the primary comb lines.** Without the primary comb lines (A), the AC trace shows no temporal fringes. When the primary comb lines are present (B,C), temporal fringes with a period matching the separation of the primary comb lines are clearly observed.

## IV. Numerical simulations

In our model we numerically studied the Kerr comb generation using formalisms of either the nonlinear coupled-mode equations, for the flexibility, or the Lugiato-Lefever equation, for its computing efficiency. For the results shown in Figure 3, the Lugiato-Lefever equation is solved for an efficient modeling of ± 256 modes around the cw pump. For the results shown in this section, the nonlinear coupled-mode equations are solved for up to ± 60 modes around the cw pump, limited by the availability of computing power. In all cases, the simulation was interrupted once the solution reaches its steady state.

In the numerical simulation, we present the spectrum of the resonator as $2\pi(\nu_j - \nu_{j0})/\gamma_0 = \nu_{FSR}(j - j_0) + \delta\nu_{j,j_0}$, where $\nu_j = \omega_j/2\pi$ is the linear frequency of the mode, $2\gamma_0$ is the FWHM of the pumped mode, $\nu_{FSR}$ is the dimensionless local averaged free spectral range of the resonator (in the simplest case of no mode interaction it is $2\nu_{FSR} = (\nu_{j_0+1} - \nu_{j_0-1})/(\gamma_0/2\pi)$), and $\delta\nu_{j,j_0}$ is the dimensionless GVD parameter. For the microresonator used in this study, $\nu_{FSR} = 1283.965$ and $\gamma_0 = 2\pi \cdot 90 MHz$.). Figure S13 plots the dimensionless GVD parameter as a function of mode number. In the simulation shown in Figure 3, the experimentally measured resonant frequencies, whenever possible, and $Q$-factors of the fundamental mode family are input directly into the model. For modes beyond our measurement capability, we assume the GVD is normal without higher order dispersions and local dispersion disruptions induced by modal interactions. Namely, $\delta\nu_{j,j_0} \cong -\frac{D_2}{2}(j - j_0)^2$. The procedure is justified by the good

S-14

agreements between the COMSOL calculations and the dispersion measurements (Figures 1b, S5a, and S5b) and the small TOD from the COMSOL calculation.

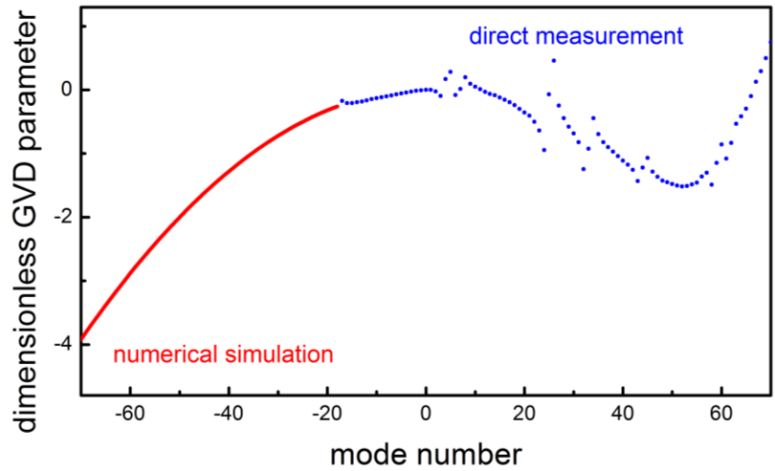

**Figure S13 | Dimensionless GVD parameters used in numerical modeling.** In the simulation shown in Figure 3, the experimentally measured resonant frequencies, whenever possible, and $Q$-factors of the fundamental mode family are input directly into the model (blue curve and datapoints). For wavelength range not covered by the measurement, we assume the GVD is normal without higher order dispersions and local dispersion disruptions induced by modal interactions (red curve).

Below we also show numerical simulations results when only the second order dispersion and attenuation were considered. These scenarios allow us to better and more rapidly understand the properties of the comb generation in normal GVD resonators. In the first simulation effort, we found that the broad phase locked Kerr comb exists in the microresonator having a normal GVD and no higher order dispersions. Furthermore, the $Q$-factor is assumed to be a constant across the whole wavelength range. The generated pulse has a very specific shape and it corresponds to a high-order dark pulse (or a manifold of dark pulses) travelling inside the resonator (Fig. S14).

To demonstrate the impact of the wavelength-dependent $Q$-factors of the resonator modes on the mode-locking, we solved the same problem with the introduction of resonance linewidth in the forms of $\gamma_j = \gamma_{j0}[1 + 0.003(j - j_0)^2]$ and $\gamma_j = \gamma_{j0}[1 + 0.01(j - j_0)^2]$. As the result, the spectral shape of the comb profile as well as the pulse shape changed drastically (Fig. S15). This simulation shows the importance of the wavelength-dependent $Q$-factors in stabilizing and shaping the pulse structures.

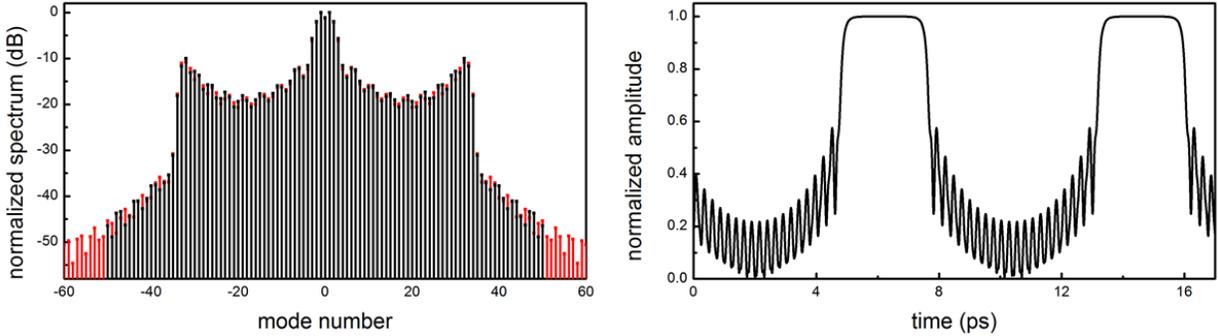

**Figure S14 | Kerr comb generated in a microresonator characterized by a large normal GVD and a wavelength independent $Q$-factors.** In this simulation, we assume the microresonator has no higher-order dispersions and its GVD is characterized by $D_2 = 0.03$. Furthermore, the $Q$-factor is assumed to be a constant across the whole wavelength range. The pump power is 49 times larger than the threshold and the resonance red-detuning is $17.4\gamma_0$.

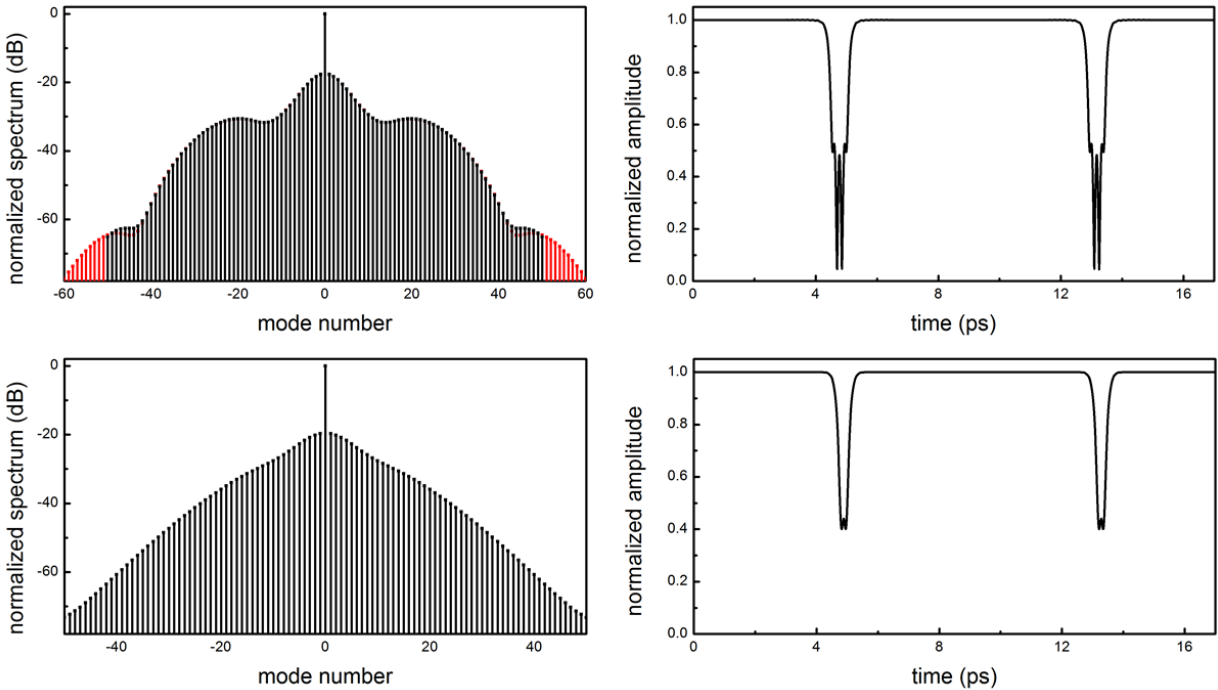

**Figure S15 | Kerr comb generated in a microresonator characterized by a large normal GVD and a wavelength dependent $Q$-factors.** Different from Figure S14, here we assume the microresonator has a wavelength-dependent $Q$-factor and its resonance linewidth is in the forms

of $\gamma_j = \gamma_{j0}[1 + 0.003(j - j_0)^2]$ (top) and $\gamma_j = \gamma_{j0}[1 + 0.01(j - j_0)^2]$ (bottom). The resonance red-detuning is $14.2\gamma_0$ and $11.5\gamma_0$, respectively.

To characterize the numerical artifact due to limited number of modes taken into consideration, we repeated the simulations for 121 modes. We observed that the solution (comb spectra, pulse width and shape) only has a relatively weak dependence on the number of modes when the modes are more than 100 in the simulations. Furthermore, the artifact was mainly observed on the spectral wings. For the modes close to the carrier, the comb line intensities vary only by roughly 1% between simulations with 101 and 121 modes.

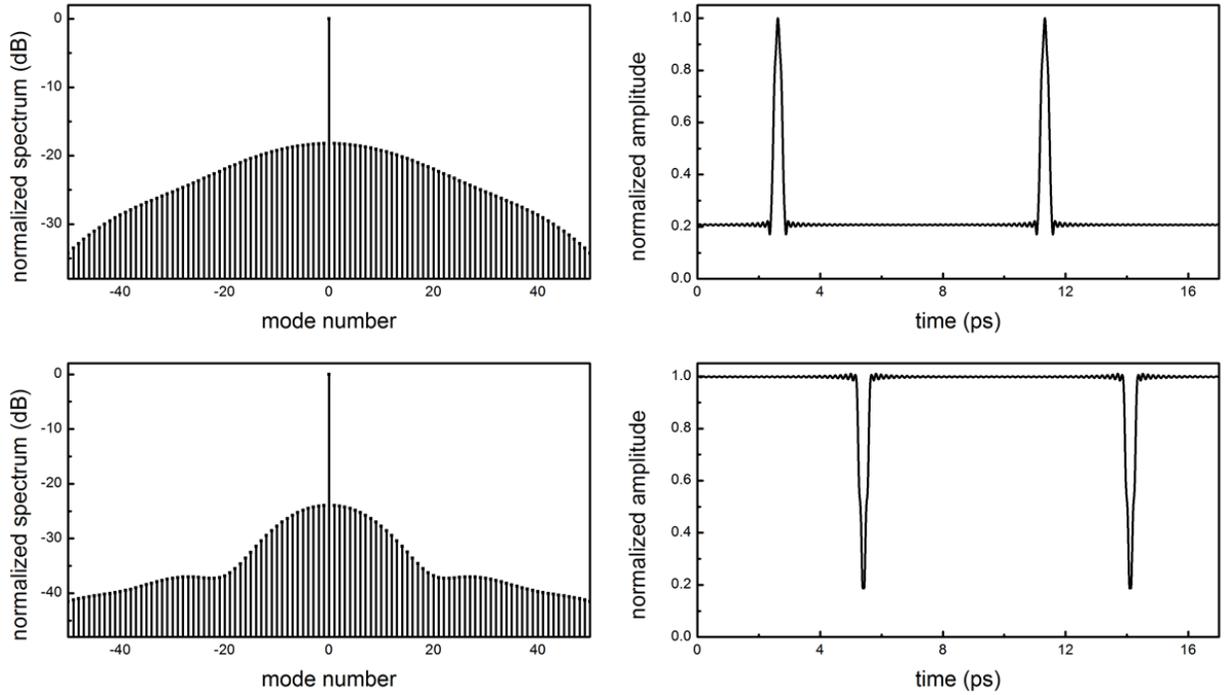

**Figure S16 | Kerr comb generated in a microresonator characterized by a small normal GVD and a wavelength dependent $Q$-factors.** For microresonators possessing a small normal GVD, both bright pulse (top) and dark pulse (bottom) can be generated. For the bright pulse generation shown here, $D_2 = 0.003$ and $\gamma_j = \gamma_{j0}[1 + 0.003(j - j_0)^2]$. The pump power is 49 times larger than the threshold and the resonance red-detuning is $10\gamma_0$. For the dark pulse generation shown here, $D_2 = 0.002$ and $\gamma_j = \gamma_{j0}[1 + 0.001(j - j_0)^2]$. The pump power is 25 times larger than the threshold and the resonance red-detuning is $7.2\gamma_0$.

Now we reduced the GVD value and repeated the simulation. As the result, a possibility of both bright and dark pulse generation was found. The number of attractors corresponding to generation of stable mode-locked pulses increased significantly as compared with the one for the case of larger GVD. Examples of the Kerr combs found the simulations are shown in Figure S16. Different from the case of large normal dispersion where only dark pulses exist, both bright and dark pulses are possible depending on the exact combination of dispersion and bandpass filter bandwidth. Experimentally, the mode-mismatched coupling also plays a role in changing the pulse shape as the imperfect coupling [S14] acts as an external filtering. A microresonator with add-drop ports will serve as a better platform for further investigation on the dark solitons [S15].

There exist multiple other solutions besides the fundamentally mode locked frequency combs generating short pulses. Dynamical solutions, such as breathers, are available. Multi-pulse regimes are feasible. Sometimes multiple pulses overlap, creating unexpected pulse shapes. For example, it is possible to generate square pulses directly out of the microresonator (Figure 3c). The simulation shows that tuning the profile of the $Q$-factors as well as the GVD is a powerful way to significantly increase the capability of these microresonators to generate arbitrary optical pulse shape.

### V. Analytic solution of normal-dispersion Kerr frequency comb

Here we look for the Gaussian solution of Eq. (1) located at cw background and use the variational method to find parameters of the solution [S16].

$$\begin{cases} A(T,t) = A_c + A_p(T,t) \\ A_c = \sqrt{P_c}e^{i\varphi_c} \\ A_p(T,t) = \sqrt{\dfrac{P_p}{\sqrt{\pi}}}\left[\exp\left(\dfrac{t}{\sqrt{2}\tau}\right)^2\right]^{-1-iq} e^{i\varphi_p} \end{cases} \tag{S1}$$

where $P_c$ is the power of the cw background, $\varphi_c$ is the phase of the background wave, $P_p$ is the pulse peak power ($E_p = P_p\tau$ is the pulse energy), $q$ is the chirp, $\tau$ is the pulse duration, and $\varphi_p$ is the phase of the pulse.

Substituting Eq. (S1) into Eq. (1) and assuming that the pulse energy is much lower than the cw energy but the pulse peak power is much higher than the DC background ($P_c T_R/P_p\tau \gg 1$ and $P_p/P_c \gg 1$), we can get the equation describing the cw background as

$$\sqrt{P_c}\left(\alpha + \dfrac{T_c}{2} + i\delta_0 - i\gamma P_c\right) = i\sqrt{T_c P_{in}}e^{i(\varphi_{in}-\varphi_c)} \tag{S2}$$

and the approximate solution is

$$\begin{cases} \varphi_{in} - \varphi_c \cong \dfrac{\alpha + T_c/c}{\delta_0} \\ P_c \cong \dfrac{T_c P_{in}}{\delta_0^2}\left(1 + \dfrac{2T_c\gamma P_{in}}{\delta_0^3}\right) \end{cases} \quad (S3)$$

On the other hand, the time-dependent part of Eq. (1) can be written as

$$\begin{cases} T_R \dfrac{\partial}{\partial T} A_p + \dfrac{i}{2}\beta_{2\Sigma}\dfrac{\partial^2}{\partial t^2} A_p - i\gamma|A_p|^2 A_p = R(T,t) \\ R(T,t) = \dfrac{T_c}{2\Omega_f^2}\dfrac{\partial^2}{\partial t^2} A_p - \left(\alpha + \dfrac{T_c}{2} + i\delta_0\right) A_p + \\ \quad i\left[\gamma\left(|A_c + A_p|^2(A_c + A_p) - |A_p|^2 A_p\right) - \dfrac{\gamma}{T_R}\int\limits_{-T_R/2}^{T_R/2} A|A|^2 dt\right] \end{cases} \quad (S4)$$

To describe the behavior of the pulse generated in the resonator we have to find values of four parameters: $P_p$, $\varphi_p$, $q$, and $\tau$. The parameters are connected by a set of self-consistent equations which can be found using variational approach [S16]. We introduce the Lagrangian density $\mathcal{L} = \dfrac{T_R}{2}\left(A_p^* \dfrac{\partial A_p}{\partial T} - A_p \dfrac{\partial A_p^*}{\partial T}\right) - \dfrac{i}{2}\left(\beta_{2\Sigma}\left|\dfrac{\partial}{\partial t} A_p\right|^2 + \gamma|A_p|^4\right)$ and the variation of the Lagrangian density results in the unperturbed nonlinear Schrödinger equation

$$\dfrac{\delta\mathcal{L}}{\delta A^*} = \dfrac{\partial\mathcal{L}}{\partial A^*} - \dfrac{\partial}{\partial T}\dfrac{\partial\mathcal{L}}{\partial(\partial A^*/\partial T)} - \dfrac{\partial}{\partial t}\dfrac{\partial\mathcal{L}}{\partial(\partial A^*/\partial T)} = $$
$$T_R \dfrac{\partial}{\partial T} A_p + \dfrac{i}{2}\beta_{2\Sigma}\dfrac{\partial^2}{\partial t^2} A_p - i\gamma|A_p|^2 A_p = 0 \quad (S5)$$

Taking into account that $A$ does not depend on $T$ directly, we write

$$\dfrac{\partial}{\partial T} A_p = \dfrac{\partial A_p}{\partial P_p}\dfrac{\partial P_p}{\partial T} + \dfrac{\partial A_p}{\partial \varphi_p}\dfrac{\partial \varphi_p}{\partial T} + \dfrac{\partial A_p}{\partial q}\dfrac{\partial q}{\partial T} + \dfrac{\partial A_p}{\partial \tau}\dfrac{\partial \tau}{\partial T} \quad (S6)$$

From Eqs. (S1), (S5), and (S6), we can write the Lagrangian of the system and the Lagrangian equations as

$$L = -i\dfrac{\beta_{2\Sigma}P_p}{4\tau}(1 + q^2) - \dfrac{i}{2\sqrt{2\pi}}\gamma P_p^2 \tau + $$
$$\dfrac{i}{4} P_p T_R \left[2q\dfrac{\partial \tau}{\partial T} - \tau\left(\dfrac{\partial q}{\partial T} - 4\dfrac{\partial \varphi_p}{\partial T}\right)\right] \quad (S7)$$

$$\dfrac{d}{dT}\left(\dfrac{\partial L}{\partial \dot{r}_j}\right) - \dfrac{\partial L}{\partial r_j} = \int\limits_{-\infty}^{\infty}\left(R^*\dfrac{\partial A_p}{\partial r_j} - R\dfrac{\partial A_p^*}{\partial r_j}\right) dt \quad (S8)$$

where $\dot{r}_j = \{\partial P_p/\partial T, \partial \varphi_p/\partial T, \partial q/\partial T, \partial \tau/\partial T\}$ and $r_j = \{P_p, \varphi_p, q, \tau\}$.

Again, under the assumption that the pulse energy is much lower than the cw energy but the pulse peak power is much higher than the DC background ($P_c T_R/P_p \tau \gg 1$ and $P_p/P_c \gg 1$), we can get the equations describing the Gaussian pulse as

$$\begin{cases} T_R \dfrac{dE_p}{dT} = -E_p \left[ T_c + 2\alpha + T_c \dfrac{1+q^2}{2\Omega_f^2 \tau^2} + \dfrac{2\sqrt{2}}{(\pi(9+q^2))^{1/4}} \gamma \sqrt{P_p P_c} \sin(\varphi_c - \varphi_p - \varphi_q) \right] \\[6pt] T_R \dfrac{d\varphi_p}{dT} = \dfrac{\beta_{2\Sigma}}{2\tau^2} + \dfrac{5}{4\sqrt{2\pi}} \gamma P_p - \delta_0 - \dfrac{qT_c}{2\Omega_f^2 \tau^2} \\[6pt] T_R \dfrac{dq}{dT} = -\dfrac{T_R}{E_p} q \dfrac{dE_p}{dT} + \dfrac{\beta_{2\Sigma}}{\tau^2}(1+q^2) + \dfrac{1}{\sqrt{2\pi}} \gamma P_p - \left( T_c + 2\alpha + \dfrac{3}{2} T_c \dfrac{1+q^2}{\Omega_f^2 \tau^2} \right) q \\[6pt] T_R \dfrac{d\tau}{dT} = -\dfrac{T_R}{2E_p} \tau^2 \dfrac{dE_p}{dT} + \beta_{2\Sigma} q - T_c \dfrac{3q^2-1}{4\Omega_f^2} - \dfrac{\tau^2}{2}(T_c + 2\alpha) \\[6pt] \sqrt{3-iq} = (9+q^2)^{1/4} e^{i\varphi_q} \end{cases} \quad (S9)$$

Further assuming that $q^2 \gg \Omega_f^2 \tau^2 \gg 1$, we finally reach the approximate solution

$$\begin{cases} E_p \cong \dfrac{8\sqrt{10\pi}}{15} \dfrac{\beta_{2\Sigma}^{\frac{3}{2}} \Omega_f^2 \sqrt{\delta_0}}{T_c \gamma} \\[6pt] \sin(\varphi_c - \varphi_p - \varphi_q) \cong -\dfrac{9}{64\sqrt{5}} \dfrac{(1+q^2)(2(9+q^2))^{1/4} T_c^3 \sqrt{\delta_0}}{\beta_{2\Sigma}^3 \Omega_f^6 \sqrt{\gamma P_c}} \\[6pt] q \cong \dfrac{4\beta_{2\Sigma} \Omega_f^2}{3T_c} \\[6pt] \tau \cong \dfrac{2\sqrt{5}}{3} \dfrac{\beta_{2\Sigma}^{\frac{3}{2}} \Omega_f^2}{T_c \sqrt{\delta_0}} . \end{cases} \quad (S10)$$

**Supplementary References:**